# ANALYTICAL STUDY OF PRE-CONGESTION NOTIFICATION (PCN) TECHNIQUES

Marwah Almasri[1], Khaled Elleithy[2], and Abdul Razaque[3]

[1]Department of Computer Science and Engineering, University of Bridgeport, Bridgeport, CT 06604
maalmasr@bridgeport.edu

[2]Department of Computer Science and Engineering, University of Bridgeport, Bridgeport, CT 06604
elleithy@bridgeport.edu

[3]Department of Computer Science and Engineering, University of Bridgeport, Bridgeport, CT 06604
arazaque@bridgeport.edu

## ABSTRACT

*Maintaining the quality of service (QOS) and controlling the network congestion are quite complicated tasks. They cause degrading the performance of the network, and disturbing the continuous communication process. To overcome these issues, one step towards this dilemma has been taken in form of Pre-congestion notification (PCN) technique. PCN uses a packet marking technique within a PCN domain over IP networks. It is notified by egress node that works as guard at entry point of network. Egress node gives feedback to communicating servers whether rate on the link is exceeded than configured admissible threshold or within the limit. Based on this feedback, admission decisions are taken to determine whether to allow/block new coming flows or terminate already accepted. The actual question is about selection of right algorithm for PCN domain. In this paper, we investigate the analytical behavior of some known PCN algorithms. We make slide modifications in originality of PCN algorithms without disquieting working process in order to employ those within similar types of scenarios. Our goal is to simulate them either in highly congested or less congested realistic scenarios. On the basis of simulation done in ns2, we are able to recommend each PCN algorithm for specific conditions. Finally, we develop a benchmark that helps researchers and scientific communities to pick the right algorithm. Furthermore, the benchmark is designed to achieve specific objectives according to the users' requirements without congesting the network.*

## KEYWORDS

*Pre-congestion notification (PCN) technique, Random Early Detection (RED), Explicit Congestion Notification (ECN), Token bucket (TB), Bandwidth Metering (BM), Additional Buffer Technique (AB).*

## 1. INTRODUCTION

With the revolution of technology, the numerous users of the Internet face many challenging issues that highly affect the quality of service (QoS). However, the Internet Engineering Task Force (IETF) has come up with the idea of pre-congestion notification (PCN) in order to avoid congestion of highly loaded network to assure the quality of service (QoS) within a Diffserv domain [1]. Excessive network load causes packet loss in the network. Furthermore, PCN maximizes the use of recourses over the link. PCN has three types of nodes, which are ingress, interior, and egress nodes. To avoid congestion, PCN uses admission control mechanism (AC) to limit the number of flows, and flow termination (FT) mechanism to remove some already accepted flows. The network's load is measured inside the PCN domain and packets are marked according to the load condition [6].

Many algorithms have been introduced to avoid the congestion and to measure the network's load. Active queue management mechanisms have positive impacts on the performance of the Internet [2]. Random early detection (RED) algorithm is one of the active queue management mechanisms to avoid congestion [2]. It is deployed in Internet [2]. RED decreases the number of dropped packets in routers. Also, it reduces the delay especially in interactive services by providing smaller average queue size. Another benefit of using this kind of algorithm is to ensure availability of the buffer for all arriving packets to control the lock-out behavior [2]. Explicit congestion notification (ECN) mechanism is another approach to avoid unnecessary dropped packets and delay in low-bandwidth TCP connections [13]. Token bucket (TB) and bandwidth metering (BM) algorithms measure the network load and signal the PCN egress node in case of congestion [14]. These both later algorithms are varied in their algorithmic complexities. Token bucket algorithm is basically a bit counter, which is updated only when a packet arrives. As a result, token bucket algorithm has a limited complexity; whereas the bandwidth metering algorithm needs more memory requirements to store the arriving packets. Bandwidth metering (BM) algorithm is helpful especially in highly congested networks [14]. The objective of this paper is to determine the weakness and strength of five algorithms supporting to PCN domain.

The paper is structured as follows. Section 2 discusses the related work regarding the PCN and the various techniques used for measuring the network load. Section 3 shows an overview of the PCN architecture and summaries existing well known techniques. Section 4 describes the simulation setup. Section 5 provides analysis of simulation results and section 6 discusses the results. Finally, section 7 concludes the paper on the basis of findings.

## 2. RELATED WORK

In this section, we discuss some salient features of PCN techniques. New bandwidth measurement technique is introduced in [3]. The paper discusses about based admission control algorithm and the performance of PCN on VBR video services. Also, it studies and analyzes the benefit of using PCN mechanism. An additional buffering technique is proposed and implemented using NS-2 based simulator. Authors validate the findings on the basis of simulation. 26.5% network utilization has been increased through this technique [4]. Different admission control (AC) methods in PCN have been investigated [5]. Authors discuss over admission flows in PCN based on excess traffic. The paper discusses that marking is occurred owing to weak pre-congestion signals. It also studies the performance of probe-based AC (PBAC) and congestion-level estimate based AC (CLEBAC) through simulation and mathematical modeling to deploy the results. Furthermore, it is observed that it is more influential in challenging conditions such as on/off traffic, low traffic aggregation, and delayed media [5].

The paper [6] discusses the encoding in IPv4 header through PCN marking mechanism. It concludes the difference between existing approaches, and suggests the enhancement of PCN design. Many algorithms have been proposed to configure the PCN threshold rate in single and dual marking PCN domain [7]. Authors suggest more requirements to incorporate in single marking than dual marking. The paper discusses that dual marking has higher resource efficiency as compared with a single marking in PCN architecture [7]. Two-layer architecture that uses various types of algorithms is introduced in [8]. The paper [15] introduces a new congestion control algorithm called modified forward active network congestion control algorithm (MFACC), which uses RED algorithm to control the queue length of the router and avoid packets loss. It also studies the active detection and the passive indication mechanism. It concludes on basis of simulation that MFACC algorithm resolves many problems found in previous used techniques, enhances the quality of service (QoS), and reduces the delay and loss packets rate [15]. RED algorithm is used with responsive and non-responsive flows [16]. The goal of this paper is to study the network performance and efficiency using NS-2 simulator. The simulations results prove that RED algorithm is working effectively even with non-responsive flows. It is also proven that RED reduces the lock out phenomenon and delay that causes of increasing the throughput [16].

We do a comprehensive analytical study of five existing algorithms using NS-2 simulator and plot the strengths and weaknesses of each technique. These techniques are: random early detection (RED), explicit congestion notification (ECN), token bucket (TB), bandwidth metering (BM), and an additional buffer technique (AB). On basis of findings, we make benchmark that helps to understand the depth of each technique in PCN domain. To validate these strengths and weaknesses of these algorithms, the realistic scenarios have been built to measure the behavior of each that supports to make the benchmark. Finally, we recommend each algorithm in specific conditions to achieve more targets..

## 3. AN OVERVIEW OF PCN ARCHITECTURE AND EXISTING WELL KNOWN TECHNIQUE

In this section, we demonstrate the design of PCN and present many used techniques inside the PCN domain that have tremendous impact on avoiding the congestion in the network.

Pre-congestion notification (PCN) has gained a lot of attention especially for the needs of emerging technology. The PCN uses two main mechanisms, which are admission control (AC) and flow termination (FT). The admission control is used to determine whether to accept or block new flows. On the other hand, the flow termination is used to terminate some of the already accepted flows [1]. Within the PCN domain, there are three types of nodes as illustrated in Figure 1: ingress node, interior node, and egress node. The ingress node and the egress node are located at the boundaries of the PCN network [3]. The interior node is responsible for measuring the congestion level inside the PCN domain by calculating the congestion level estimator, (CLE) that uses an exponential weighted moving average (EWMA). The value of (CLE) falls between one and zero. After computing the (CLE) value, the egress node sends signal to the ingress node in order to decide whether to accept or block new flows [4]. When the CLE is 1, there is a pre-congestion while 0 means there is no congestion. CLE can be calculated as follows:

$$CLE_n = Thr * (1 - CLE_W) + CLE_W * CLE_{n-1} \qquad (1)$$

$CLE_W$ donates the CLE weight. *Thr* is a threshold value that can be either 1 or 0 depending on the packet marking status [3].

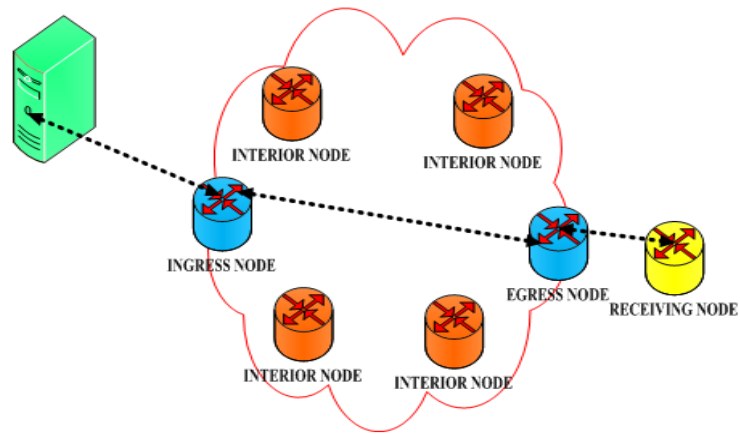

Figure 1: Describes the PCN architecture which consists of three types of nodes: ingress, egress, and interior nodes.

In addition, PCN has defined two rate thresholds, an admissible and a supportable rate threshold (Ar), (Sr) which provide three types of pre-congestion. Figure 2, summaries these types as follows: when the PCN traffic rate r < Ar, there is no congestion in the network. Hence, new flows are accepted. In contrast, when the PCN traffic rate r > Ar, the link 'L' in the network is Ar-pre-congested and the overload value is said to be Ar-overload. As a result, new flows are not accepted. Another issue when the PCN traffic rate r > Sr, the link 'L' in the network is Sr-pre-congested and the overload value is said to be Sr-overload. Consequently, we need to reduce the value of rate r(L) by terminating already accepted flows [8].

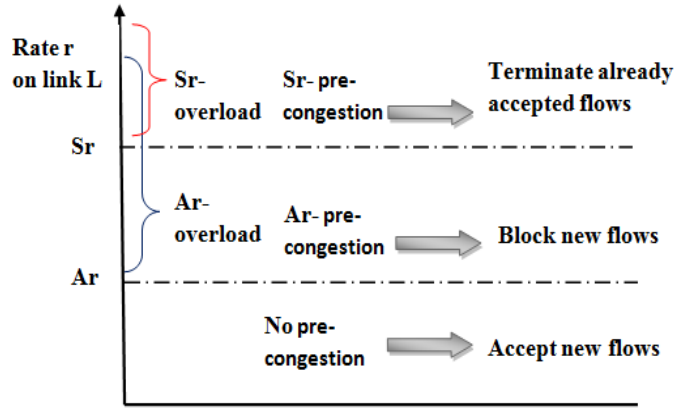

Figure 2: Describes the three types of pre-congestion.

## 3.1 Random Early Detection (RED)

Random early detection (RED) algorithm is used to avoid the congestion in the network especially in the PCN mechanism. It detects early congestion in order to avoid the congestion and to enhance the TCP throughput performance. RED algorithm's idea is basically based on the buffering queue length [7]. It computes an average queue size (*avg*) by an exponential weighted moving average. Then it compares this average (*avg*) with two other parameters which are a minimum threshold (*min_thr*) and a maximum threshold (*max_thr*). If average queue size (*avg*) falls below the minimum threshold (*min_thr*), then no packets are marked or dropped. On the other hand, if average queue size (*avg*) goes above the maximum threshold (*max_thr*), then the packets are marked. Another issue is when the average queue size (*avg*) is between the minimum threshold (*min_thr*) and the maximum threshold (*max_thr*) values, the packets are marked relatively to a probability $P_A$ [9]. The general model of RED algorithm is described in Algorithm 1. In addition, the packet-marking probability $P_p$ increases linearly from 0 to $Max_p$ along with the average queue size *avg* as follows:

$P_p = Max_p \ (avg - min\_thr) / (max\_thr - min\_thr)$ (2)

The final packet-marking probability $P_A$ increases along with the increment of the counter since the last marked packet [9] as follows:

$P_A = P_p / (1 - count * P_p)$ (3)

The mechanism of this algorithm comprises of two main parts. One is to compute the average queue size (*avg*), and the other to calculate the probability $P_A$ that the packets are marked with [9].

Algorithm. 1: RED algorithm

```
For each arriving packet
    Compute the average queue size avg
    If min_thr ≤ avg ≤ max_thr
        {
            Compute probability  $P_A$
            Mark packet according to $P_A$
        }
    Else if  max_thr  ≤ avg
        {
            Mark the arriving packet
        }
```

## 3.2 Explicit Congestion Notification (ECN)

Explicit congestion notification (ECN) is another technique that is used in PCN model in order to avoid congestion. It is based on random early detection (RED) algorithm.

It signals the incipient congestion to notify the TCP sender to reduce the sending rate window [10]. ECN protocol uses congestion experienced (CE), that is the code point, located at the packet header. This is useful to indicate that there is congestion rather than just dropping the packets. As a result, using ECN in TCP/IP networks has the benefit of not dropping or delaying the packets. In addition, the TCP and IP header need to be modified to hold extra bits for ECN protocol as described in Figure 3, for the IP header, and Figure 4, for the TCP header.

If the value of the code point is "10", or "01", the packets are ECN capable. If the value of the code point is "00", the packets are not ECN capable. However, if the value of the code point is "11" which is the CE code point, meaning that there is congestion and the packets are marked. Hence, when receiving the marked packet with CE code point, TCP connection should reduce its sending rate of packets [11].

| version | Header Length | DSCP Field | ECT | CE | Total Length | |
|---|---|---|---|---|---|---|
| Identification | | | | | Flags | fragment offset |
| Time to Live | | protocol | | | Header Checksum | |
| Source IP Address | | | | | | |
| Destination IP Address | | | | | | |

Figure 3: ECN in IP header.

| Source Post | | | | | | | | | | Destination Post | |
|---|---|---|---|---|---|---|---|---|---|---|---|
| Sequence Number | | | | | | | | | | | |
| Acknowledgement Number | | | | | | | | | | | |
| Header Length | Reserved Field | C W R | E C E | U R G | A C K | P S H | R S T | S Y N | F I N | | |
| TCP Checksum | | | | | | | | | | Urgent Pointer | |

Figure 4: ECN in TCP header.

## 3.3 Token Bucket (TB)

Token bucket is a measurement algorithm that is used at the interior node in the PCN domain. It's primarily goal is to limit the speed of network transmission. On the other hand, it is also helpful to determine the bandwidth usage. The token bucket is a bit counter that shows the load in the network [12], as given in Figure 5. These tokens are added to the bucket at a constant token rate (R). However, when the packets are reached at interior nodes in the PCN domain, tokens are removed from the bucket. If the aggregate bandwidth is lower than (R), the number of tokens increases. On the other hand, if the aggregate bandwidth is greater than (R), the number of tokens decreases. As a result, in the PCN architecture, packets are marked when the number of tokens is fewer than the token bucket threshold [3]. Algorithm 2, shows the algorithm for token bucket mechanism.

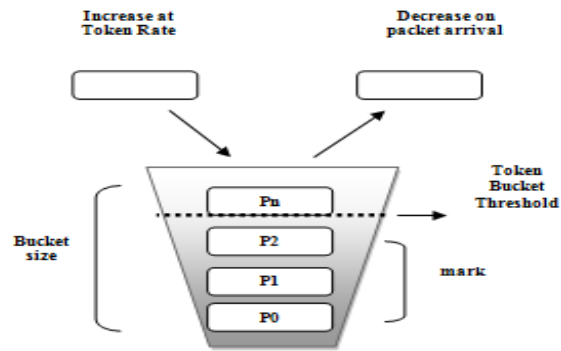

Figure 5: Token bucket mechanism.

Algorithm 2: Token bucket algorithm.

1. Bucket= empty, Token = 0;
2. *Max_rate* = maximum output rate;
3. *BC* = capacity of token bucket;
4. *L* = burst length;
5. *G* = generating rate;
6. *Max_rate* = *BC* / *L* + *G*;
7. Repeat
   If (token < *BC*)
8. accept packets
9. If (token > 0)
10. *G*= dequeue (bucket, *Max_rate*)
11. Display *Max_rate* output
12. If (t == Δt)
13. token++
End Repeat
End If
End If
End If

## 3.4 Bandwidth Metering (BM)

This mechanism is another way to measure the load in the network. It differs from the token bucket because it uses a time window technique. This algorithm marks packets when the aggregate bandwidth is greater than a predefined threshold. This algorithm is better than the token bucket because it computes the accurate measurement of the bandwidth instead of just comparing it with the token rate *(R)*. In contrast, the bandwidth metering requires more memory but its advantage outweigh this extra requirement. In addition, the base of this technique is on a sliding window with a fixed "*mi*", which is a measurement interval. During the last "*mi*", the bandwidth is measured as it receives packets and then the packets are marked if the bandwidth is higher than a bandwidth threshold [12]. Algorithm. 3, demonstrates the bandwidth metering algorithm.

Algorithm. 3:Bandwidth metering algorithm.

1. Input: (*B* = bandwidth measurement, *B_Thr* = bandwidth threshold, *mi* = fixed measurement interval)
2. Output: (*m_packet*)
3. For each packet at arrival time *t*
4. Compute *B* during the last *mi* seconds.
5. If $B(t) > B\_Thr(t)$
6. Process *m_packet*
7. End

## 3.5 An Additional Buffer Technique (AB)

In this technique, an additional buffer is used in the PCN domain as shown in Figure 6. The algorithm for this technique can be implemented as follows. The threshold rate (*Tr*) is calculated by (4) where (*Ar*) is the admissible rate threshold, and (*Or*) is the objective rate:

$Tr = Ar + Or / 2$ (4)

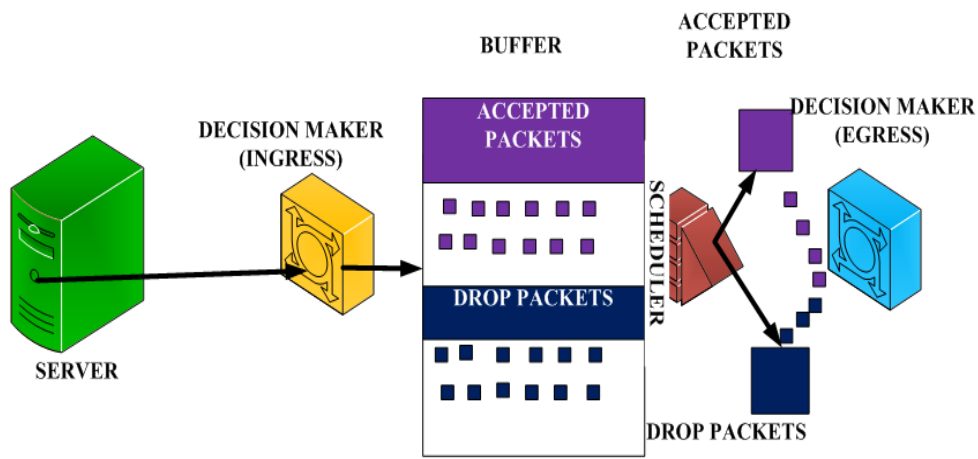

Figure 6: Describes the additional buffer technique.

The computed threshold rate (*Tr*) is important to construct the buffer [4]. All packets enter the PCN domain through the ingress node which works as the decision maker. It classifies the packets into accepted packets or dropped/marked packets and sends them to the buffer. However, if the buffer is full, incoming packets are dropped until the buffer gets some space for accepting new packets. After that, all packets including the dropped packets are sent to the scheduler. The scheduler's function is send these packets to the egress node with a priority. Accepted packets get higher priority while dropped/marked packets get lower priority. After transmitted the packets to the egress node, the packets are forwarded to the transport layer. This priority depends on the weighted value of the dropped packets (*Wd*) and the buffer (*Wb*). These two weighted values can be calculated by using (5) and (6).

$Wd = 1-(Tr/Or)$        $Wd \in [0, 1]$        (5)

$Wb = Tr/Or$        $Wb \in [0, 1]$        (6)

## 4. SIMULATION SETUP

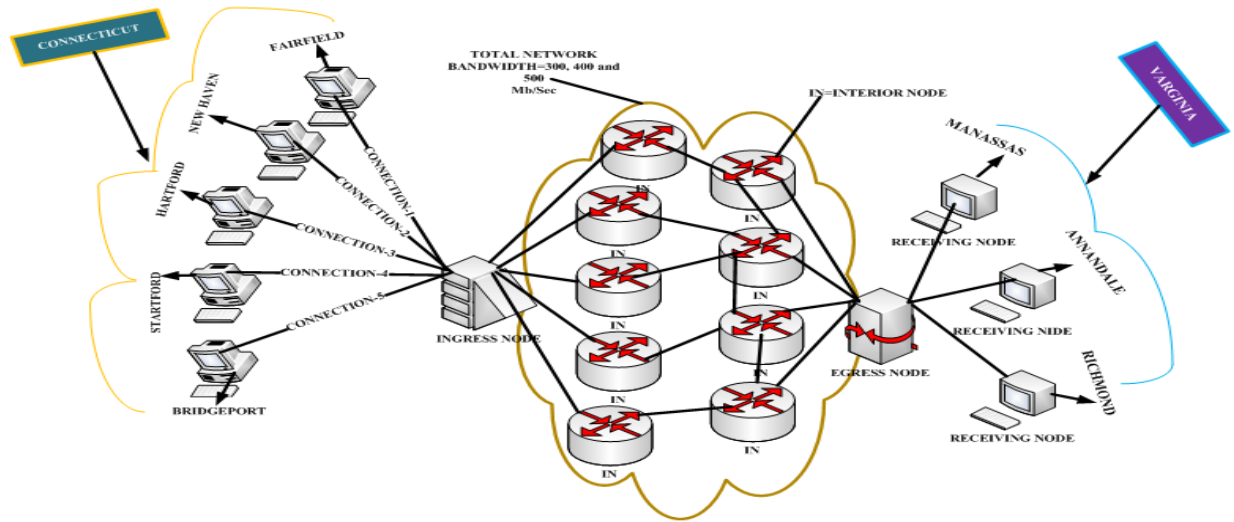

Figure 7: The simulation scenario.

Our objective is to simulate all the techniques described in section 3 based on realistic scenarios in simple and highly congested network. We make PCN domain highly congested to examine the performance of network. We assume in our simulation scenario that there are eight educational institutions located at two different states as shown in Figure 7. The objective of this scenario is to maintain the quality of service and provide better data communication among these educational institutions. On the basis of the scenario, we use parameters that help to examine the behavior of these well know techniques with different network bandwidth. These scenarios are simulated using CBR, FTP, and HTTP applications and supported with UDP and TCP layer protocols. The simulation area is 500X500 m2 and the number of mobile nodes is 50 nodes. The sensing range of the node is 250 m2. The size of packets is 1040 bytes including 40 bytes header. In our scenario, the PCN ingress node is connected via 5 links. The size of bandwidth is 300 Mbps, 400 Mbps, and 500 Mbps used in the PCN domain with these techniques. The capacity of each link is equal. If the bandwidth of the network is 300 Mbps that each link gets equal share 60 Mbps. Similarly, 400 Mbps and 500 Mbps bandwidths are divided in equal five shares. The 10 connections are established at a time and the packet generation rate is 15 packets per second. In our simulation experiment, we set 4 seconds pause time after interval of 5 minutes of simulation.

## 5. ANALYSIS OF RESULTS

### 5.1 Throughput

We use TCP as transport protocol for sending and receiving the data. The throughput in the simulation can be calculated by using (7):

$$Max\_Throughput = Buf\_Size / RTT \qquad (7)$$

Where *Max_Throughput*= Maximum throughput; *Buf_Size*= Received buffer size; *RTT*= Round Trip Time.

RED algorithm gives better throughput at the 500 Mbps bandwidth. At the same time, the token bucket algorithm has the lowest throughput rate. The average throughput is calculated 38.75 at 500 Mbps bandwidth RED algorithm whereas it is calculated 31.25 for token bucket algorithm, that makes the RED algorithm 19.4 % better than the token bucket algorithm. In addition, RED algorithm provides the best throughput at 400 Mbps bandwidth whereas token bucket and additional buffer techniques have the lowest throughput rate. The average throughput at 400 Mbps bandwidth is measured 39.25 Mbps for RED algorithm, 31.25 for token bucket technique, and 31.75 Mbps for the additional buffer technique. As a result, we have obtained 19.7 % more throughput when using RED algorithm at 400 Mbps bandwidth than using token bucket or additional buffer techniques. On the other hand, ECN algorithm produces the highest throughput at 300 Mbps bandwidth while the lowest throughput is examined at the additional buffer technique. The average throughput is calculated 34 Mbps for ECN algorithm and 29 Mbps for the additional buffer technique. Hence, ECN algorithm produces 14.7 % more throughput than the additional buffer technique. The average throughput of all techniques used in this paper is given in Figure 8, Figure 9, and Figure 10. Table 1 describes the abbreviations of different techniques used in the graphs.

Table 1: describes the abbreviations of different techniques used in the graphs.

| Abbreviation | Full Description |
|---|---|
| **TB** | Token Bucket Technique |
| **BM** | Bandwidth Metering Technique |
| **RED** | Random Early Detection Technique |
| **AB** | Additional Buffer Technique |
| **ECN** | Explicit Congestion Notification Technique |

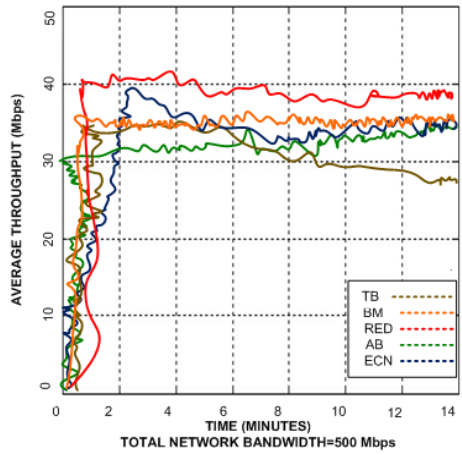

Figure 8: The average throughput at 500 Mbps bandwidth with different techniques.

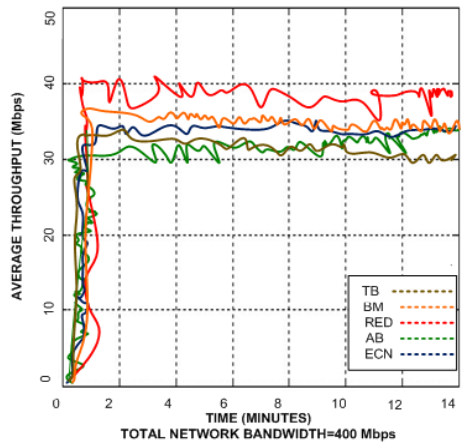

Figure 9: The average throughput at 400 Mbps bandwidth with different techniques.

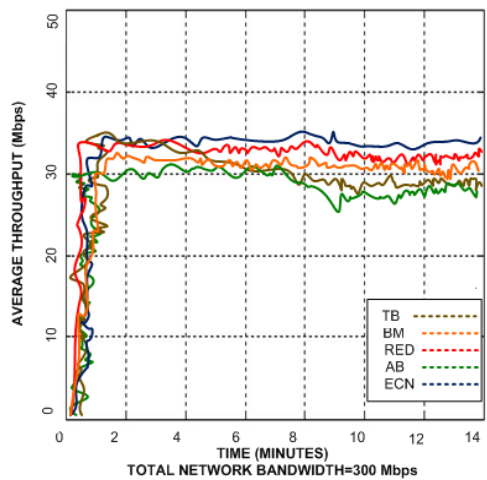

Figure 10: The average throughput at 300 Mbps bandwidth with different techniques.

### 5.2 Loss/Drop of Packets Ratio %

The loss/drop of packets ratio can be calculated by using (8) and (9) :

*LP= TSP- TAP*                                                (8)

Where LP= total number of loss packets; TSP= total number of sent packets; TAP= total number of acknowledged packets.

*DR= LP X 100 / TSP*     , Where DR = Drop rate         (9)

All techniques discussed in this paper are investigated with various network bandwidths in order to examine the packet loss ratio. Token bucket technique gives lower packet loss ratio at 500 Mbps bandwidth whereas the additional buffer technique provides higher packet loss ratio at the same bandwidth. Token bucket reduces 22.3% of average packet loss than the additional buffer technique. In contrast, at 400 Mbps bandwidth, the lowest packet loss ratio is obtained by ECN algorithm whereas the highest is examined at RED algorithm. Hence, 5.3% reduction in packet loss ratio with ECN algorithm is calculated. Conversely, RED algorithm produces lower packet loss ratio at 300 Mbps bandwidth whereas the additional buffer technique gives higher packet loss ratio. RED algorithm reduces 18.6% the average packet loss ratio than the additional buffer technique. The average packet loss ratio is described in Figure 11, Figure 12, and Figure 13.

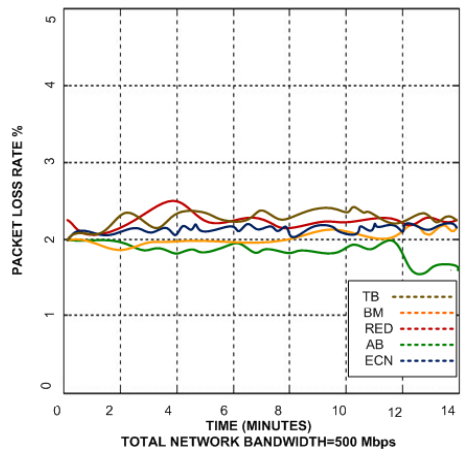

Figure 11: The average packet loss rate at 500 Mbps bandwidth with different techniques.

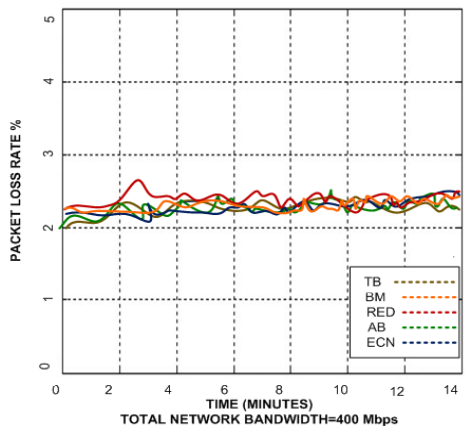

Figure 12: The average packet loss rate at 400 Mbps bandwidth with different techniques.

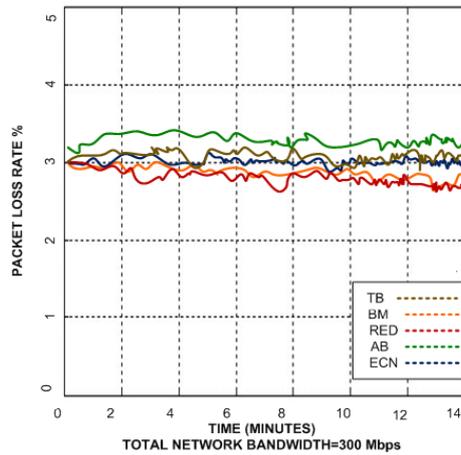

Figure 13: The average packet loss rate at 300 Mbps bandwidth with different techniques.

## 5.3 Admitted Sessions

We have set the optimal TCP window size that is most useful to control the congestion even in interior nodes. We apply the following formula to find the optimal window size:

*Optimal window size= 2 X B X DP*     (10)
Where *B*= Bandwidth; *DP*= Delay of Product.

The optimal window size helps to determine the current size of window in order to establish the session as per bandwidth capacity of network. The additional buffer technique has 71 admitted session whereas token bucket technique has 64 admitted sessions at 500 Mbps bandwidth. Hence, the additional buffer technique has admitted 9.86% more than the token bucket technique. Furthermore, the additional buffer technique and the bandwidth metering techniques admitted 63 sessions which are measured as 12.7 % more sessions than the token bucket technique at 400 Mbps bandwidth. In addition, RED and bandwidth metering algorithms accept 58 sessions comparing to 53 secessions through token bucket technique at 300 Mbps bandwidth. Thus, 8.6 % more sessions are gained by RED and bandwidth metering algorithms than the token bucket technique. The averages of sessions admitted are shown in Figure 14, Figure 15, and Figure 16.

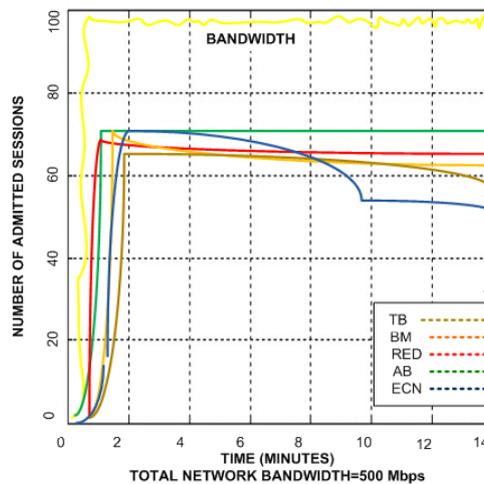

Figure 14: The average number of sessions admitted at 500 Mbps bandwidth with different techniques.

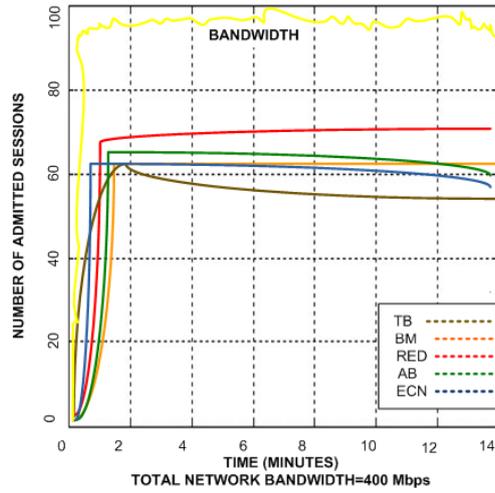

Figure 15: The average number of sessions admitted at 400 Mbps bandwidth with different techniques.

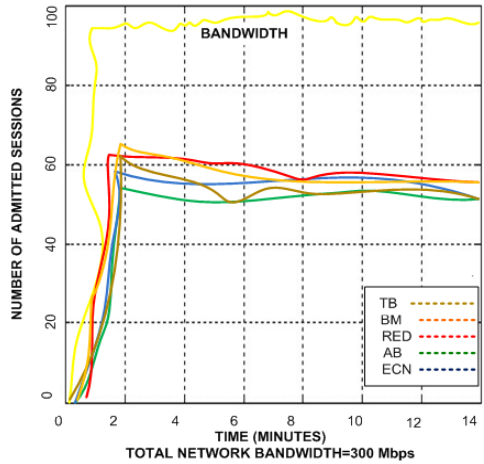

Figure 16: The average number of sessions admitted at 300 Mbps bandwidth with different techniques.

## 6. DISCUSSION OF RESULTS

The main objective of this work is to determine the most effective technique in PCN domain. We consider three important parameters to justify the weaknesses and the strengths of each algorithm. The first is an average throughput, the second is packet loss/drop ratio, and third is number of average admitted sessions. As the total network bandwidth increases, the RED algorithm is a better choice to enhance the throughput because of the small average queue size, which causes high throughput and low average delay. In contrast, as the total network bandwidth decreases, the ECN algorithm is a better choice to enhance the throughput due to the less sensitivity to network parameters, which also improves throughput and reduces delay. However, token bucket and additional buffer techniques always have the lower throughput. Regarding the packet loss rate, RED algorithm is found to be the best as the network bandwidth decreases and Token bucket technique is more reliable as the network bandwidth increases. RED algorithm marks packets randomly with a certain probability $P_A$ instead of discarding them when average queue size *avg* is between the values of *min_thr* and *max_thr*. As a result, it encourages the early detection of congestion and thus adjusts the window size to avoid packet loss. Moreover, the additional buffer technique enables to admit more sessions as the bandwidth increases and RED algorithm enables to admit more sessions as the bandwidth decreases. However, the Token bucket technique always admits lower number of sessions comparing with other techniques. The additional buffer technique admits more sessions because it reduces the admissible rate threshold (*Tr*) when the network is congested, so, it increases the number of admitted sessions. To conclude, RED algorithm is the best technique that reflects more positive results than others. RED has a higher average throughput and a lower delay time. It provides a fair service of packet dropping due to the better loss packet rate and admits a satisfactory number of sessions. The benchmark of our results is shown in Table 2.

Table 2: Benchmark of all techniques used.

| FACTORS | 300 Mbps | | 400 Mbps | | 500 Mbps | |
|---|---|---|---|---|---|---|
| AVERAGE THROUGHPUT | **TECHNIQUE** | **AVG** | **TECHNIQUE** | **AVG** | **TECHNIQUE** | **AVG** |
| | **AB** | 29 | **AB** | 31.75 | **AB** | 33.5 |
| | **ECN** | 34 | **ECN** | 33.75 | **ECN** | 34.5 |
| | **TB** | 30.25 | **TB** | 31.25 | **TB** | 31.25 |
| | **BM** | 31 | **BM** | 35.25 | **BM** | 35.5 |
| | **RED** | 33 | **RED** | 39.5 | **RED** | 38.75 |
| AVERAGE PACKET LOSS RATE | **TECHNIQUE** | **AVG** | **TECHNIQUE** | **AVG** | **TECHNIQUE** | **AVG** |
| | **AB** | 3.38 | **AB** | 2.4 | **AB** | 1.85 |
| | **ECN** | 3 | **ECN** | 2.35 | **ECN** | 2.18 |
| | **TB** | 3.1 | **TB** | 2.38 | **TB** | 2.38 |
| | **BM** | 2.95 | **BM** | 2.38 | **BM** | 2.3 |
| | **RED** | 2.75 | **RED** | 2.48 | **RED** | 2.23 |
| AVERAGE ADMITTED SESSIONS | **TECHNIQUE** | **AVG** | **TECHNIQUE** | **AVG** | **TECHNIQUE** | **AVG** |
| | **AB** | 55 | **AB** | 63 | **AB** | 71 |
| | **ECN** | 56 | **ECN** | 62 | **ECN** | 66 |
| | **TB** | 53 | **TB** | 55 | **TB** | 64 |
| | **BM** | 58 | **BM** | 63 | **BM** | 65 |
| | **RED** | 58 | **RED** | 62 | **RED** | 65 |

## 7. CONCLUSION

In this paper, we have reported the results of simulation results and investigated the performance of five different PCN techniques with different bandwidth. These five PCN techniques are implicitly discussed with slide modification in their existing algorithms. This modification helps to determine the exact behavior of each technique in congested PCN domain. The goal of the research is to determine which technique is the most suitable in particular scenario. RED algorithm provides better performance comapred to other techniques in terms of throughput due to its small average queue size especially when there is enough bandwidth. RED has a lower packet loss ratio. It has an ability to mark packets according to probability $P_A$ instead of discarding them in small bandwidth. The number of admitted sessions are calculated more using the additional buffer technique in high bandwidth and maintains minimum sessions in case of low bandwidth whereas RED technique makes more sessions in small bandwidth.

In future work, we plan to integrate the best features of all techniques and introduce new technique in PCN domain to avoid congestion, and maintain high quality of service by achieving maximum throughput.

# REFERENCES

[1] P. Eardley, "Pre-Congestion Notification (PCN) Architecture," RFC 5559 (Informational), Jun. 2009. [Online]. Available: http://www.ietf.org/rfc/rfc5559.txt

[2] B. Braden et al. RFC2309: Recommendations on Queue Management and Congestion Avoidance in the Internet, Apr. 1998.

[3] S. Latr´e, B. De Vleeschauwer, W. Van de Meerssche, F. De Turck, P. Demeester "Design and Configuration of PCN Based Admission Control in Multimedia Aggregation Network", in *Proceedings of IEEE Globalcom*, 2009.

[4] K. Roobroeck, S. Latr´e, T. Wauters, F. De Turck, "Optimized network utilisation through buffering in PCN enabled multimedia access networks," Integrated Network Management (IM), 2011 *IFIP/IEEE International Symposium* on , vol., no., pp.1243-1249, 23-27 May 2011.

[5] M. Menth, F. Lehrieder, "Performance of PCN-Based Admission Control Under Challenging Conditions," Networking, *IEEE/ACM Transactions* on , vol.20, no.2, pp.422-435, April 2012.

[6] M. Menth et al., "A Survey of PCN-Based Admission Control and Flow Termination," *IEEE Communications Surveys & Tutorials*, vol. 12, no. 3, 2010.

[7] M. Menth, M. Hartman, "Threshold configuration and routing optimization for PCN-based resilient admission control", *Journal of Computer Networks*, 2009.

[8] M. Menth, F. Lehrieder, "Performance Evaluation of PCN-Based Admission Control", in *Proceedings of Quality of Service*, 2008. IWQoS 2008. 16th International Workshop on , 2008, pp 110-120.

[9] S. Floyd and V. Jacobson. Random Early Detection Gateways for Congestion Avoidance. *IEEE/ACM Transactions* on Networking, 1(4):397–413, Aug. 1993.

[10] K. Ramakrishnan, S. Floyd, and D. Black. RFC3168: The Addition of Explicit Congestion Notification (ECN) to IP, Sept. 2001.

[11] R. Arora, O. Yang, H. Zhang, "TCP Networks with ECN over AQM". *OPNETWORK* 2002, August 2002, Washington DC, USA.

[12] S. Latre, B. De Vleeschauwer, W. Van de Meerssche, S. Perrault, F. De Turck, P. Demeester, K. De Schepper, C. Hublet, W. Rogiest, S. Custers, and W. Van Leekwijck, "An Autonomic PCN based Admission Control Mechanism for Video Services in Access Networks," in *IEEE ACNM*, 2009.

[13] P. S, Floyd, "TCP and explicit congestion notification", in ACM SIGCOMM Computer Communication Review, October 1994.

[14] K.Venkateswara Rao, E.Hemalatha, "Performance enhancement by measuring traffic at edge routers in a high speed networks", in *International Journal of Engineering Science and Technology*, vol. 3, no. 6, June, 2011.

[15] J. Wang, X. Wang, H. Wang, M. Huang, L. Ma, Z. Pan , "The Research Of Active Network Congestion Control Algorithm Based On Operational Data," *Communications and Networking in China, 2007. CHINACOM '07. Second International Conference* on , vol., no., pp.213-217, 22-24 Aug. 2007.

[16] A. Krempa, "Analysis of RED algorithm with responsive and non-responsive flows", *Poznan University of Technology Academic Journals,* 2007.

**Authors**

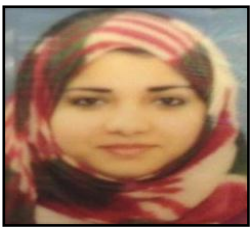

**Mrs. Marwah Almasri**: is a Ph.D. student in the Computer Science & Engineering Department at the University of Bridgeport. She received her MBA in Management Information System (MIS) from the University of Scranton, PA, in 2011. She received award from MIS department at the University of Scranton for her outstanding work. She holds a bachelor degree in Computer Science & Engineering from Taibah University in Medina, Saudi Arabia. Her research interests include congestion mechanisms, wireless and sensor networks, computer networks, mobile computing, and network security.

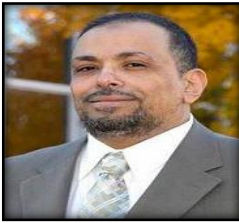

**Dr. Khaled Elleithy**: is the Associate Dean for Graduate Studies in the School of Engineering at the University of Bridgeport. His research interests are in the areas of, network security, mobile wireless communications formal approaches for design and verification and Mobile collaborative learning. He has published more than one two hudereds research papers in international journals and conferences in his areas of expertise.

Dr. Elleithy is the co-chair of International Joint Conferences on Computer, Information, and Systems Sciences, and Engineering (CISSE).CISSE is the first Engineering/ Computing and Systems Research E-Conference in the world to be completely conducted online in real-time via the internet and was successfully running for four years. Dr. Elleithy is the editor or co-editor of 10 books published by Springer for advances on Innovations and Advanced Techniques in Systems, Computing Sciences and Software.

Dr. Elleithy received the B.Sc. degree in computer science and automatic control from Alexandria University in 1983, the MS Degree in computer networks from the same university in 1986, and the MS and Ph.D. degrees in computer science from The Center for Advanced Computer Studies in the University of Louisiana at Lafayette in 1988 and 1990, respectively. He received the award of "Distinguished Professor of the Year", University of Bridgeport, during the academic year 2006-2007.

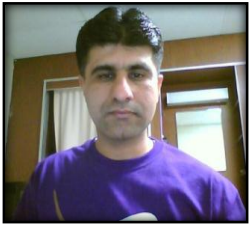

**Mr. Abdul Razaque** is PhD student of computer science and Engineering department in University of Bridgeport. His current research interests include the design and development of learning environment to support the pedagogical activities in open, large scale and heterogamous environments, collaborative discovery learning and the development of mobile applications to support mobile collaborative learning (MCL), the congestion mechanism of transmission of control protocol including various existing variants, delivery of multimedia applications. He has published over 40 research contributions in refereed conferences, international journals and books. He has also presented his work more than 10 countries. During the last two years he has been working as a program committee member in IEEE, IET, ICCAIE, ICOS, ISIEA and Mosharka International conference.

Abdul Razaque is member of the IEEE, ACM and Springer. Abdul Razaque served as Assistant Professor at federal Directorate of Education, Islamabad. He completed his Bachelor and Master degrees in computer science from university of Sind in 2002. He obtained another Master degree in computer Science with specialization of multimedia and communication (MC) from Mohammed Ali Jinnah University, Pakistan in 2008. Abdul Razaque has been directly involved in design and development of mobile applications to support learning environments to meet pedagogical needs of schools, colleges, universities and various organizations.